\documentclass[openbib,twocolumn]{revtex4}
\usepackage{graphicx}
\usepackage{dcolumn}
\usepackage{amsmath}
\usepackage{epsfig}

\begin{document}
\newlength{\bibindent}
\setlength{\bibindent}{0pt}
\bibliographystyle{unsrt}

\textbf{Valanju \emph{et. al.} (VWV) reply:} 
The central argument of Pendry and Smith (PS) \cite{PS} is:
``\emph{VWV correctly calculate positively refracted modulation fronts 
at a vacuum/NIM interface, but wrongly interpret their normal as the 
direction of group velocity $\vec{v}_{g}$. 
Both the phase velocity $\vec{v}_{p}$ and $\vec{v}_{g}$ 
refract negatively so the angle between them is $\theta_{pg}=\pi$.
The propagation of the front is crabwise,
and antiparallel to $\vec{v}_{p}$, as required by Veselago.}'' 
However, PS do not show how $\theta_{pg}=\pi$, 
or Veselago's strictly monochromatic refraction \cite{Ves}, 
can yield ``crabwise'' waves, which are nothing but 
the rigorously derived inhomogeneous waves of VWV \cite{nim1}
with $\theta_{pg} \ne \pi$.

PS's use of Eq.2 to imply $\theta_{pg}=\pi$ is incorrect because it neglects 
the non-collinearity of refracted $\vec{k}(\omega)$ 
at different $\omega$ due to dispersion. 
PS do not derive the value of $\theta_{pg}$, and their assertion, 
that $\vec{v}_{g}$ must be either parallel or antiparallel to $\vec{k}$
in an isotropic NIM, is wrong because there is no unique $\vec{k}(\omega)$ 
after refraction at even an isotropic, but dispersive, NIM.
The direction of $\vec{k}(\omega)$ can not be both independent of $\omega$
as in their Eq.2, and yet depend on $\omega$ as required to generate inhomogeneous
waves in their Figures 2 and 3. In fact, these figures agree with 
the wave inhomogeneity first shown in VWV, but not with PS's Eq.2 
or Veselago's rays which imply unphysical, negatively refracted homogeneous 
wavefronts. PS associate $\vec{v}_{g}$ with the Poynting vector $\vec{P}$, 
but VWV showed that the NIM dispersion creates 
inhomogeneous waves at a PIM-NIM interface because
$\vec{v}_{g}$ and $\vec{v}_{p}$ (and consequently $\vec{P}$) 
refract in different ($\pm$) directions.

Fig.3 of PS is underived: collinear $\vec{k}$'s in Eq.2 cannot give
wave inhomogeneity ($\theta_{pg} \ne \pi$). 
In contrast, by adding transverse $\pm k$ components to each $k(\omega)$ in 
Eq.1 of Ref.\cite{nim1}, we derive our Fig.1. It
shows that static, transverse spatial features do not
qualitatively change VWV's results. 
The wave modulation fronts are still oriented in the positive direction, 
the waves are still inhomogeneous, and hence decay rapidly. 
Static effects of transverse $\pm k$ are unaffected by 
causality because they carry no time-varying signal. 
Hence they were correctly omitted from VWV to present with clarity 
the first theory of causal signal refraction by NIM.

PS's conclusion, that Veselago was correct in saying that
$\vec{v}_{g}$ is antiparallel to $\vec{v}_{p}$ ($\theta_{pg}=\pi$), 
disagrees with the inhomogeneous waves in their Figures 2 and 3. 
Veselago considered only monochromatic refraction, which by definition contains 
no modulation to misalign with $\vec{v}_{p}$. 
Neither PS's Eq.2 nor Veselago's monochromatic ray diagrams represent
inhomogeneous waves with $\theta_{pg} \ne \pi$ because they neglect dispersion. 

Although all equations in Ref.\cite{nim1} are correct,
the inadvertent use of ``wave'' for ``wavefront''
in the title, and the association of $\vec{v}_{g}$ with energy flux
in the first paragraph of Ref.\cite{nim1}, may have caused the PS Comment.
However, neither the central result (wave inhomogenization),
nor any equations of Ref.\cite{nim1}, are disproved by PS.
The key issue here is wave inhomogenization that invalidates ray diagrams, 
not the direction of $\vec{v}_{g}$. Causality implies that, after 
refraction,  $\vec{v}_{g}$ differs from $\vec{v}_{p}$ and energy velocity 
$\vec{v}_{E}$, causing anomalously rapid, dispersive wave decay \cite{nim2}.

In summary, PS's neglect of dispersion at the interface 
(by \emph{assuming} a unique $\vec{k}(\omega)$) makes
their Eq.2 wrong, irrelevant, and inconsistent with their
underived Figures 2 and 3. Instead, their figures support 
VWV's derivation of wave inhomogenization by oblique refraction in NIM.

\vskip 0.2in
\begin{flushleft}
     P.M. Valanju, R.M. Walser, and A.P. Valanju

     The University of Texas at Austin.
\end{flushleft}

\begin{figure}
 \epsfig{file=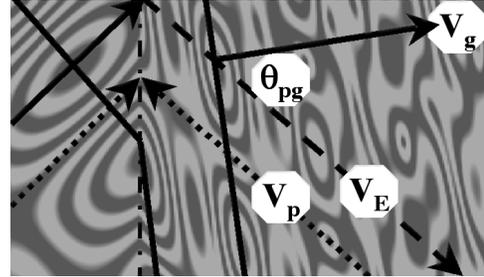,width=2.5 in}
  \caption
   {Intensity of wave with three $\omega$ components, 
    each with two transverse $\pm k(\omega)$.
    Transversely chopped modulation fronts (solid) still
    refract positively in NIM. 
    They are not normal to the $\vec{v}_{p}$ (dotted) 
    or Poynting ($\vec{P}$ or $\vec{v}_{E}$) directions (dashed),
    i.e, $\theta_{pg} \ne \pi$. 
    The resulting inhomogeneous waves decay anomalously fast,
    even for infinitesimal Im[$n_{p}(\omega)$].
   }
  \label{fig}
\end{figure}


\end{document}